# Rapid online deep artifact suppression for real-time spiral bSSFP CMR with blipped-CAIPI simultaneous multi-slice imaging at 1.5 T

Julius Åkesson[1,2,3], Iulius Dragonu[4], Einar Heiberg[2,3], Tina Yao[1], Rebecca Baker[1], Ruta Virsinskaite[1], Daniel Knight[1], Vivek Muthurangu[1], Jennifer Steeden[1]

[1]: *Centre for Translational Cardiovascular Imaging, Institute of Cardiovascular Science, University College London, London, United Kingdom.*

[2]: *Clinical Physiology, Department of Clinical Sciences Lund, Lund University, Skåne University Hospital, Lund, Sweden.*

[3]: *Department of Biomedical Engineering, Faculty of Engineering, Lund University, Lund, Sweden.*

[4]: *Research & Collaborations GBI, Siemens Healthcare Ltd, Camberley, United Kingdom.*

*: Correspondence: *jennifer.steeden@ucl.ac.uk*

## ABSTRACT

*Purpose:* Real-time (RT) bSSFP MRI enables fast free-breathing cardiovascular imaging but requires 10-16 slices for functional assessment, resulting in prolonged scan times. Simultaneous multi-slice (SMS) imaging can reduce acquisition time but when combined with non-Cartesian trajectories, it relies on iterative reconstructions that preclude online use. This study investigates deep artifact suppression to facilitate rapid, online reconstruction of RT-SMS.

*Methods:* A spiral bSSFP SMS RT sequence with two simultaneously acquired slices was implemented at 1.5 T. Reconstruction used slice separation in k-space, followed by deep artifact suppression in image space using a 3D U-Net. Ten healthy volunteers were imaged. RT-SMS image quality and reconstruction time were compared between deep artifact suppression and compressed sensing (CS) reconstructions. Left (LV) and right (RV) ventricular volumes at end diastole (EDV) and end systole (ESV) and LV mass (LVM) were compared between RT-SMS with deep artifact suppression and reference-standard breath-hold (BH) imaging.

*Results:* The RT-SMS acquisition was ~13× faster than BH imaging (15 s vs 3 min 15 s). RT-SMS reconstruction using deep artifact suppression was ~50× faster than CS (30 s vs 24 min 55 s). Deep artifact suppression consistently outperformed CS in quantitative and qualitative image quality ($p<0.001$). Functional agreement between BH and RT-SMS with deep artifact suppression was good (LVEDV: -7.5±6.8 ml, LVESV: -0.9±4.2 ml, RVEDV: -6.4±8.4 ml, RVESV: 0.2±10.7 ml, LVM: -10.3±11.0 g).

*Conclusion:* Online deep artifact suppression reconstruction for RT-SMS bSSFP CMR enables free-breathing short-axis coverage with a substantial reduction in acquisition and reconstruction time while maintaining diagnostic image quality.

## INTRODUCTION

Cardiovascular magnetic resonance (CMR) is the clinical reference-standard for evaluating ventricular volumes and function. Conventionally, multi-slice, electrocardiogram (ECG)-gated, balanced steady state free precession (bSSFP), Cartesian cine imaging is used to measure volumes with 1-2 slices acquired per breath-hold. This is time-consuming (often > 5 minutes) and difficult for patients who struggle with breath-holding[1]. Alternatively, real-time imaging can be used to enable rapid, continuous acquisition during free-breathing[1]. In this approach, total scan times are primarily determined by the number of slices required for full ventricular coverage (~10-16 slices).

A solution to this problem is simultaneous multi-slice (SMS) imaging, where more than one slice (S > 1) is acquired at the same time[2], resulting in a scan time reduction proportional to the SMS factor (S). Simultaneously acquired slices can be subsequently separated using controlled aliasing methods[3], e.g. blipped-controlled aliasing in parallel imaging (blipped-CAIPI[4]), which applies gradient blips to impart slice-dependent phase modulation.

Although current state-of-the-art SMS techniques use Cartesian trajectories, non-Cartesian trajectories are desirable to achieve real-time imaging with high spatiotemporal resolution. However, non-Cartesian SMS imaging typically requires iterative reconstruction methods[5], resulting in long reconstruction times[6]. This could be mitigated using deep learning (DL), which has been shown to enable low-latency online reconstruction through single-pass deep artifact suppression in image space[7–9].

In this study, we applied DL to non-Cartesian, real-time SMS reconstruction. The aims of this study were to: (i) Develop a real-time spiral bSSFP SMS (RT-SMS) sequence at 1.5 T for rapid free-breathing assessment of ventricular volumes; (ii) Develop a deep artifact suppression method for rapid online reconstruction (< 30 seconds latency); (iii) Compare image quality of RT-SMS data reconstructed using deep artifact suppression and state-of-the-art compressed sensing (CS); and (iv) Assess left-ventricular (LV) and right-ventricular (RV) volumes from RT-SMS with deep artifact suppression reconstruction, in comparison to reference-standard breath-hold (BH) cine imaging.

## METHODS

### *Pulse Sequence*

An in-house real-time spiral bSSFP blipped-CAIPI SMS-2 sequence (with two simultaneously acquired slices, as shown in Figure 1A) was developed on a 1.5 T Siemens MAGNETOM Aera system (Siemens Healthineers, Forchheim, Germany). Multi-band excitation was implemented by modifying a single-band sinc RF pulse[10], such that for a short-axis stack of $S_{tot}$ slices (where $S_{tot}$ is the total number of slices in the stack) two slices were excited simultaneously, separated by $S_{tot}/2$ slice positions (i.e. slice $i$ and slice $i+S_{tot}/2$). Real-time acquisition was achieved using a variable-density spiral trajectory[11] (see Figure 1), with eight uniformly spaced interleaves per timeframe, and tiny golden angle[12] rotation between timeframes (≈32.04°).

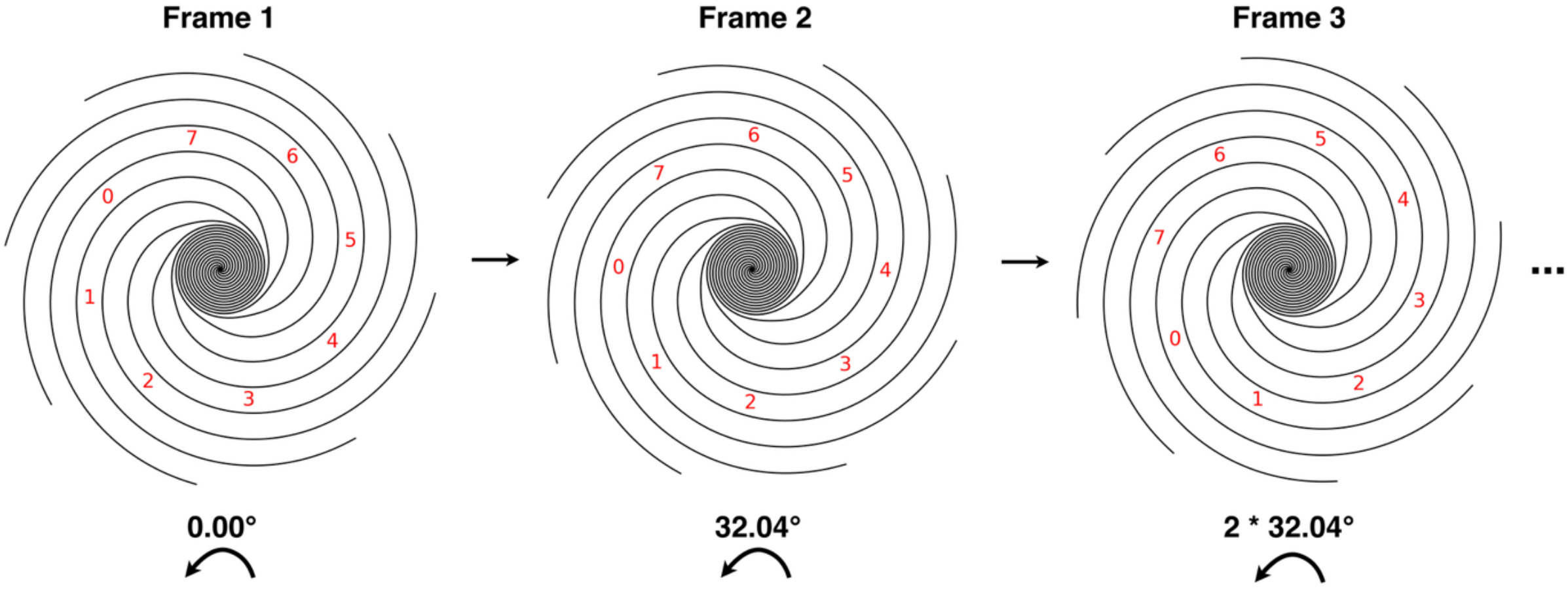

**Figure 1:** *This figure shows the variable density spiral trajectory used to achieve real-time SMS-2. The first three timeframes are shown. The inner 20% of the k-space radius is under-sampled by 1.6×, and the outer 70% is under-sampled by 14.8× (with the remaining 10% covering the transition between the two sampling densities).*

Full sequence parameters are given in Table 1:

| | **RT-SMS** | **Reference-standard BH** |
|---|---|---|
| TR/TE (ms) | 4.54 / 1.15 | 2.14 / 1.07 |
| Flip Angle | 100$^{o}$ | 69$^{o}$ |
| FOV (mm) | 400 x 400 | 384 x 372 |
| Matrix Size | 240 x 240 | 384 x 372 |
| Pixel Spacing (mm) | 1.7 x 1.7 | 1.0 x 1.0 |
| Slice Thickness (mm) | 8.0 | 8.0 |
| Spiral Interleaves per frame | 8 | - |
| Temporal resolution (ms) | 36.32 | 40.96 |
| Number of Slices | 12-16 (6-8 SMS-2 slice-pairs with 48-64 mm slice separation) | 12-16 (no slice gap) |
| Total Acquisition time (s) | 15 $\pm$ 3 | 195 $\pm$ 28 |

***Table 1:*** *This table shows the sequence parameters for the RT-SMS and reference-standard BH cine imaging sequences in the prospective study (n=10).*

Blipped-CAIPI phase modulation ($\phi = \pm \pi/2$) was applied between consecutive TRs[13] (and thus between spiral interleaves) to aid slice separation[14] (Figure 2):

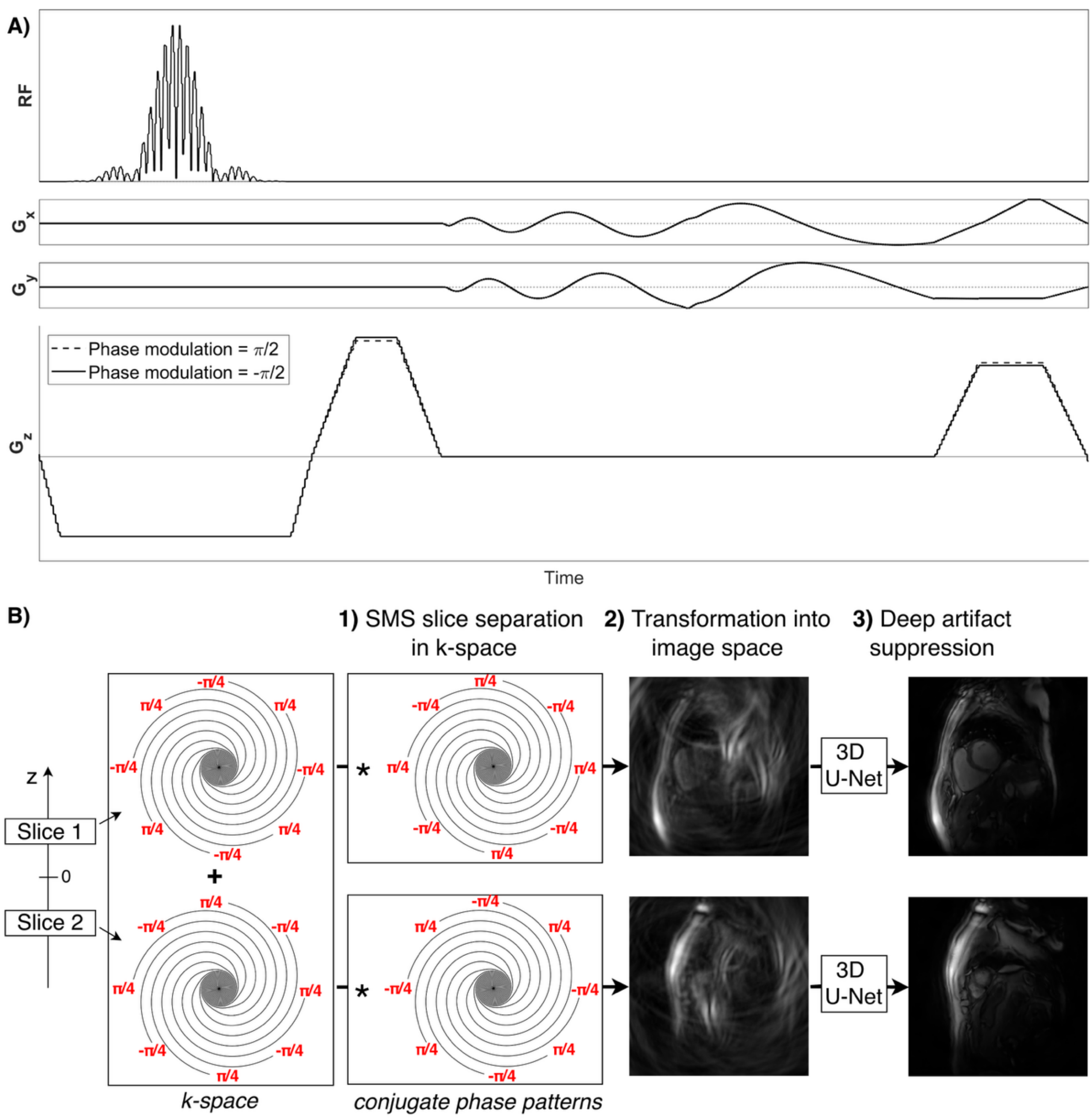

***Figure 2: A)** Pulse sequence diagram for the spiral bSSFP blipped-CAIPI SMS-2 (RT-SMS) sequence. Note: the gradient blips are integrated into the slice-select re-winder gradient (for $\phi = \pm \pi/2$). **B)** The* deep artifact suppression *reconstruction pipeline, consisting of (1): SMS slice separation in the phase-modulated SMS-2 k-space through phase de-modulation using the conjugate phase patterns; (2) transformation into image space using a NUFFT; and (3) deep artifact suppression using a 3D U-Net.*

***Reconstruction Overview***

The RT-SMS deep artifact suppression (RT-SMS-DL) reconstruction pipeline (Fig. 1B) consisted of three steps:

1) *SMS Slice Separation in k-space*

   In the acquired SMS-2 k-space data, the two superimposed slices (referred to as the primary and secondary slice) are encoded with distinct blipped-CAIPI phase modulation patterns. Therefore, slice separation can be achieved by de-modulating the acquired k-space with the conjugate of the phase pattern associated with each slice[14]. This yields two slice-specific k-space datasets for independent reconstruction (see Figure 3 for details).

2) *Transformation into Image Space*

   The separated k-spaces are each coil-compressed (using singular-value decomposition to 10 coil channels, to ensure consistent input dimensionality across datasets for the DL model) and sampling density compensation is subsequently applied[15]. This is followed by a transformation into image space using the non-uniform fast Fourier transform (NUFFT).

3) *Deep Artifact Suppression*

   Deep artifact suppression is performed in image space for each slice separately, using a 3D U-Net[16] architecture (2D + time, architecture shown in Figure 4).

The code is available online at: https://github.com/mrphys/RT-SMS-ML

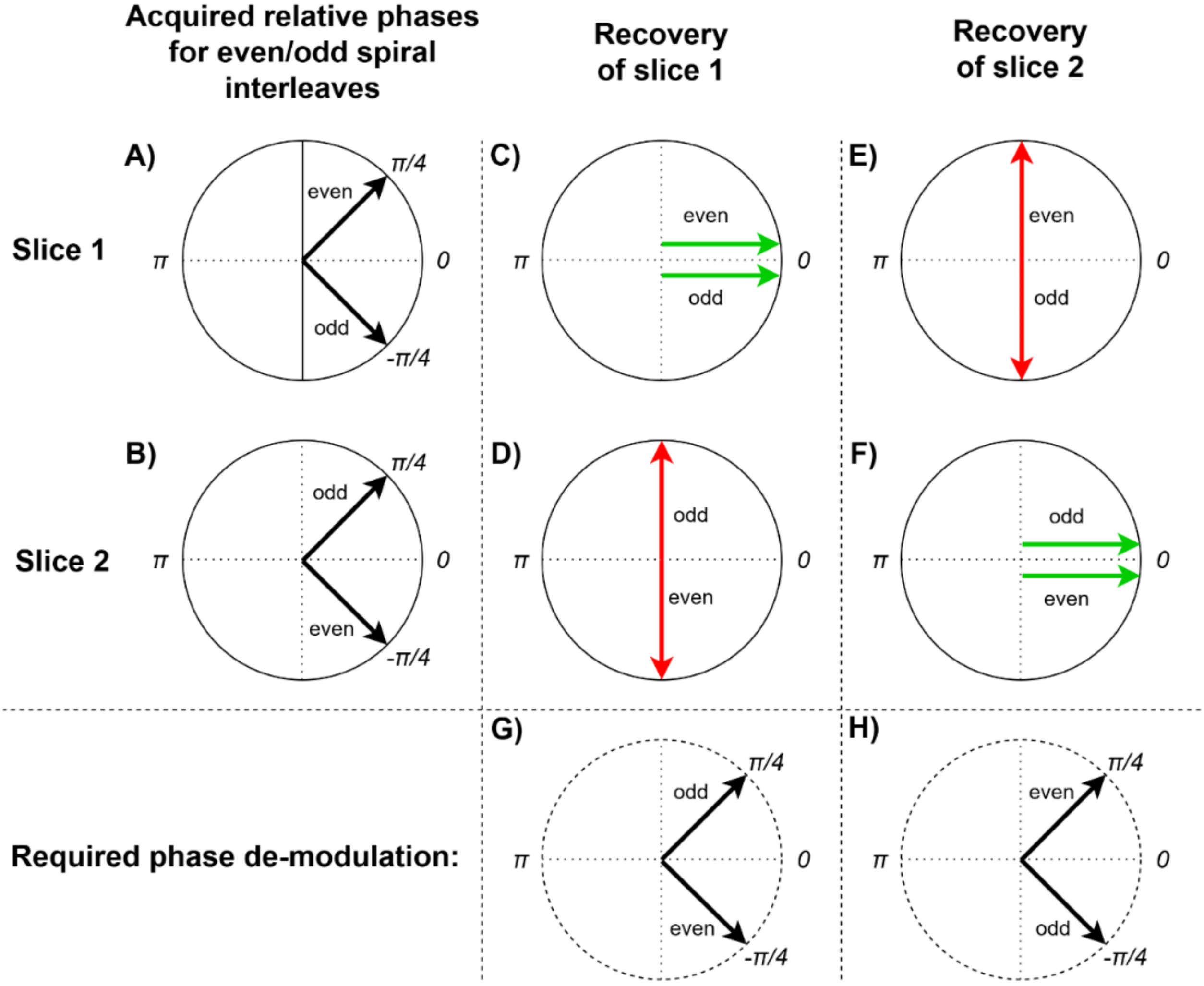

***Figure 3:*** *This figure shows how SMS slice separation is carried out in k-space by illustrating the constructive and destructive signal interference used for slice recovery in the case where the Z gradient isocenter is positioned in the center between two SMS slices (accruing phase patterns of ± π/4 between spiral interleaves). The marks "odd" and "even" refer to spiral interleaves with odd and even indices, respectively. A) the phase modulation pattern accrued during acquisition for slice 1; B) the phase modulation pattern accrued during acquisition for slice 2; C) constructively interfering phase for slice 1; D) destructively interfering phase for slice 2; E) destructively interfering phase for slice 1; F) constructively interfering phase for slice 2; G) conjugate phase de-modulation needed to recover slice 1 and suppress slice 2; H) conjugate phase de-modulation needed to recover slice 2 and suppress slice 1. Note that when the signal from one slice is recovered, the signal from the other slice is suppressed.*

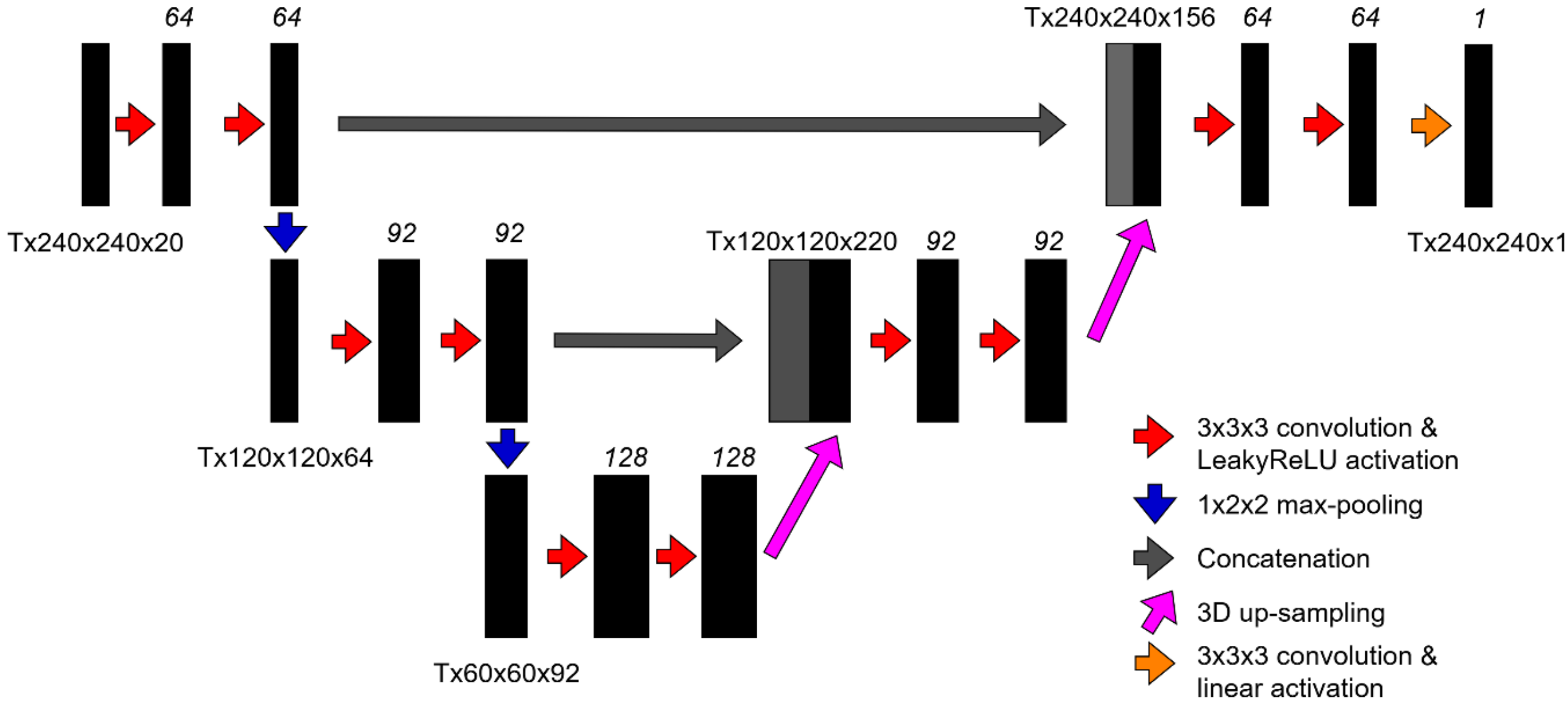

***Figure 4****: This figure shows the 3D U-Net architecture used to perform deep artifact suppression. The network had two encoding steps and two decoding steps with skip connections. Each encoding step consisted of two 3×3×3 convolutions with LeakyReLU activations followed by 1×2×2 max-pooling, and each decoding step consisted of two 3×3×3 convolutions with LeakyReLU activations followed by 3D up-sampling. The input size was T×240×240×20 and the output size was T×240×240×1, where T is the number of timeframes, 240×240 is the image matrix size, and the last dimension consists of the real and imaginary parts of the 10 coil images, compressed to a single magnitude image in the output.*

### *Model Training*

The U-Net was trained using an Adam optimizer[17] with a structural similarity index measure (SSIM) loss, on an NVIDIA RTX A6000 GPU. Synthetic paired training data were created from retrospectively gated Cartesian breath-hold bSSFP multi-slice short-axis (SAX) cine k-space data from two populations: (i) 64 subjects collected at the Royal Free Hospital, London, United Kingdom (Ethics ref. 21/EE/0037, 17/LO/1499); and (ii) 11 subjects from the publicly available Open-Access Multi-Coil k-Space Dataset for Cardiovascular Magnetic Resonance Imaging (OCMR) dataset (all subjects with fully sampled, multi-slice SAX data)[18]. The data was split into 59/6/10 subjects for training/validation/testing, with all OCMR data placed in the training set. All datasets were pre-processed by interpolation in image space to a 240×240 matrix, with 24 timeframes.

For each subject dataset, two slices of the SAX stack were pseudo-randomly selected to represent the two slices excited by the SMS-2 sequence. RT-SMS k-space data were simulated by: (i) Resampling the fully-sampled, multi-coil Cartesian k-space from each of the two selected slices onto the variable-density spiral trajectory; (ii) Applying the slice-specific blipped-CAIPI phase-modulation to each slice's multicoil complex k-space; and (iii) Summing the phase-modulated k-spaces for each channel, to generate a composite RT-SMS k-space. This simulated RT-SMS k-space was then separated and transformed into image space (as described above) to create two multi-coil, complex images with synthesized SMS-2 artifacts. During training, each slice was treated as an independent input to the deep artifact suppression model. For each slice, the corresponding ground-truth magnitude images were generated through Fourier transformation of the fully sampled Cartesian k-space data. This process was repeated 10 times per subject (using different slice-pairs each time), resulting in 1180/120/200 cine slices for training/validation/testing.

### *Prospective In Vivo Study*

To demonstrate feasibility of the proposed technique, prospective data were acquired in 10 healthy volunteers (age: 18-51 years) at Skåne University Hospital in Lund, Sweden. Written informed consent was provided by all participants, and the collection of data was approved by the Swedish Ethical Review Authority (Dnr 2025-00334-02).

RT-SMS data were collected in the SAX view during free-breathing, with 12-16 contiguous slices (slice thickness 8 mm, no inter-slice gap), acquired as 6-8 SMS-2 slice-pairs. This corresponded to a slice separation of 48-64 mm between simultaneously acquired primary-secondary slices (i.e. primary basal slices are paired with secondary apical slices across the stack), yielding a contiguous short-axis stack. Each SMS-2 acquisition required one dummy R-R interval to reach steady state, with data collected during the following R-R interval. In addition, reference-standard BH Cartesian ECG-gated bSSFP cine SAX data were also acquired. Full sequence parameters are given in Table 1.

*Inference and online implementation*

The proposed RT-SMS-DL reconstruction was implemented for online deployment using Gadgetron[19], on an external workstation with a GPU (NVIDIA GeForce RTX 4090). For comparison, the RT-SMS data were also reconstructed using the state-of-the-art spatiotemporally constrained compressed sensing (CS) reconstruction (RT-SMS-CS) presented by Yagiz et al.[5], with default parameters. In this CS approach, the SMS-combined data are directly used as input and the reconstruction jointly resolves the simultaneously excited slices, producing two separated slice images.

***In Vivo Image Analyses***

*Quantitative and Qualitative Image Quality*

All quantitative and qualitative image quality assessments were performed on two contiguous mid-ventricular short-axis slices per subject. These represent the last slice of the primary SMS-2 stack and first slice of the secondary SMS-2 stack. These were chosen to enable assessment of slice-separation and image quality for primary and secondary SMS slices.

Quantitative image quality was measured on RT-SMS-DL, RT-SMS-CS, and reference-standard BH images, using edge sharpness, estimated signal-to-noise ratio (eSNR) and blood pool-to-myocardial contrast ratio (bmCR). Edge sharpness was computed along a line profile placed perpendicularly across the septum at end diastole, as previously described[20]. eSNR was calculated as the average pixel intensity in the LV blood pool in diastole (signal) divided by the standard deviation of the pixel intensity in a background air region (noise*)*. bmCR was calculated as the average pixel intensity in the LV blood pool divided by the average pixel intensity within the LV myocardium, at end diastole.

Qualitative image ranking was performed independently by two expert observers (RV: 3 years CMR experience, and DK: 15 years CMR experience), who were blinded to the acquisition/reconstruction type. To compare RT-SMS image quality from deep artifact suppression and CS reconstructions, RT-SMS-DL and RT-SMS-CS image series were presented side-by-side (as movies) in random order, and the reviewer selected the reconstruction with preferred image quality (with the option of a tie). In addition, RT-SMS-DL was compared to reference-standard BH imaging, by presenting these image series individually (as movies), and scoring their diagnostic quality, using a five-step Likert scale with options: 1: *Non-diagnostic*, 2: *Poor*, 3: *Adequate*, 4: *Good,* and 5: *Excellent*.

*Volumetric Assessment*

LV blood pool, RV blood pool and LV myocardial segmentations were performed at end systole (ES) and end diastole (ED) from RT-SMS-DL and reference-standard BH images. This enabled calculation of LV and RV end-diastolic volume (EDV), end-systolic volume (ESV), stroke volume (SV), and ejection fraction (EF), as well as LV mass (LVM). Segmentations were performed using a pretrained nnUNet model[21], with manual correction by an expert observer (VM: 25 years of experience) using Roundel[22].

### *Statistics*

Statistical analyses were performed using SciPy (version 1.10.1) in Python, with a p-value $<0.05$ considered statistically significant. Continuous variables are reported as mean ± standard deviation. Comparisons of quantitative image scoring (edge sharpness, eSNR and bmCR) across all three groups were performed using one-way repeated measures analysis of variance (ANOVA) with post hoc testing using Holm correction for significant results. Qualitative image quality rankings between RT-SMS-DL and RT-SMS-CS were compared using paired preference testing, with a binomial test against a null hypothesis of equal preference ($p=0.5$). Likert-scale qualitative image quality scores for RT-SMS-DL and reference-standard BH images were compared using a Wilcoxon signed-rank test. Mean ventricular volumes, function and mass measured using RT-SMS-DL and reference-standard BH techniques were compared using a paired t-test, and agreement was assessed using Bland-Altman analysis.

## RESULTS

Training the deep artifact suppression model took ~128 hours. On the synthetic test set, the model achieved a mean SSIM of 0.92 ± 0.03.

### *Prospective In Vivo Study*

RT-SMS and reference-standard BH imaging data were successfully acquired in all subjects, with total acquisition times of 15 ± 3 s vs. 195 ± 28 s, respectively (~13× faster). At inference, the total reconstruction time per subject, for all RT-SMS slices using deep artifact suppression was 30 ± 1 s, compared with 1495 ± 25 s using CS (~50× faster).

Figure 5 shows representative RT-SMS images reconstructed using deep artifact suppression and CS. Both reconstruction approaches successfully separated the SMS slices. However, the corresponding X-T plots (Figure 5) demonstrated improved temporal fidelity with deep artifact suppression compared to CS. A comparison of the image quality between RT-SMS-DL and the reference-standard BH images is provided in Figure 6.

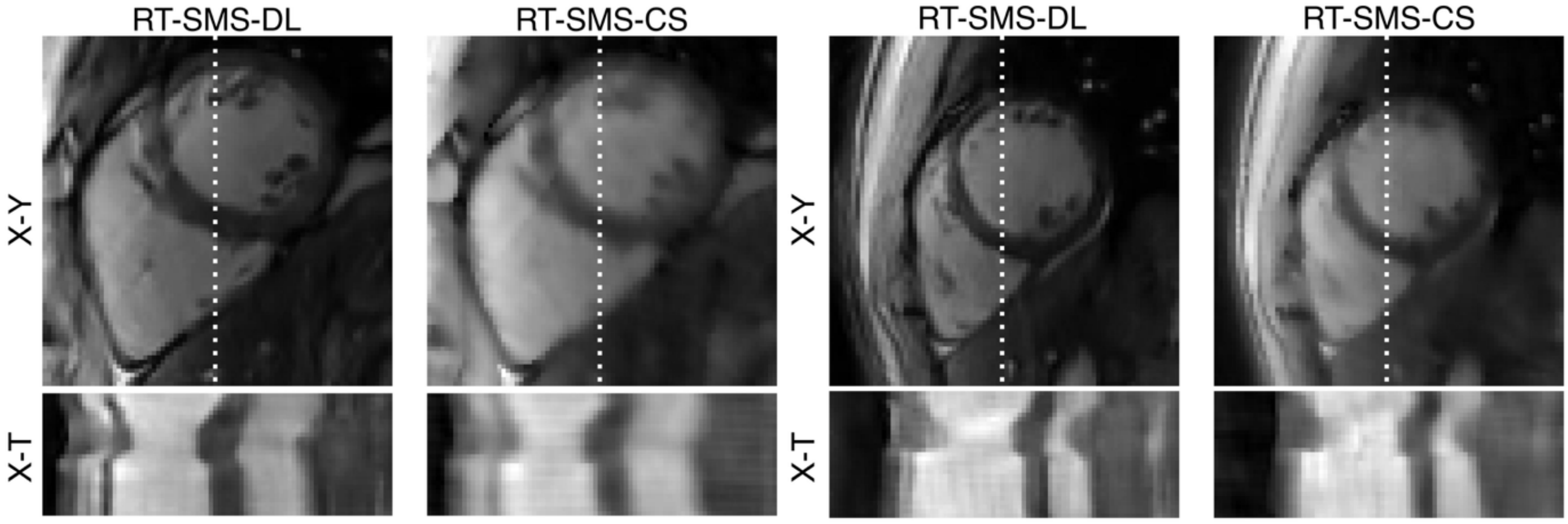

***Figure 5:*** *RT-SMS data from mid-ventricular slices in two subjects, reconstructed using* deep artifact suppression *(RT-SMS-DL) and compressed sensing (RT-SMS-CS). The X-T plots show the intensity across the vertical dotted line in the X-Y plots over time.*

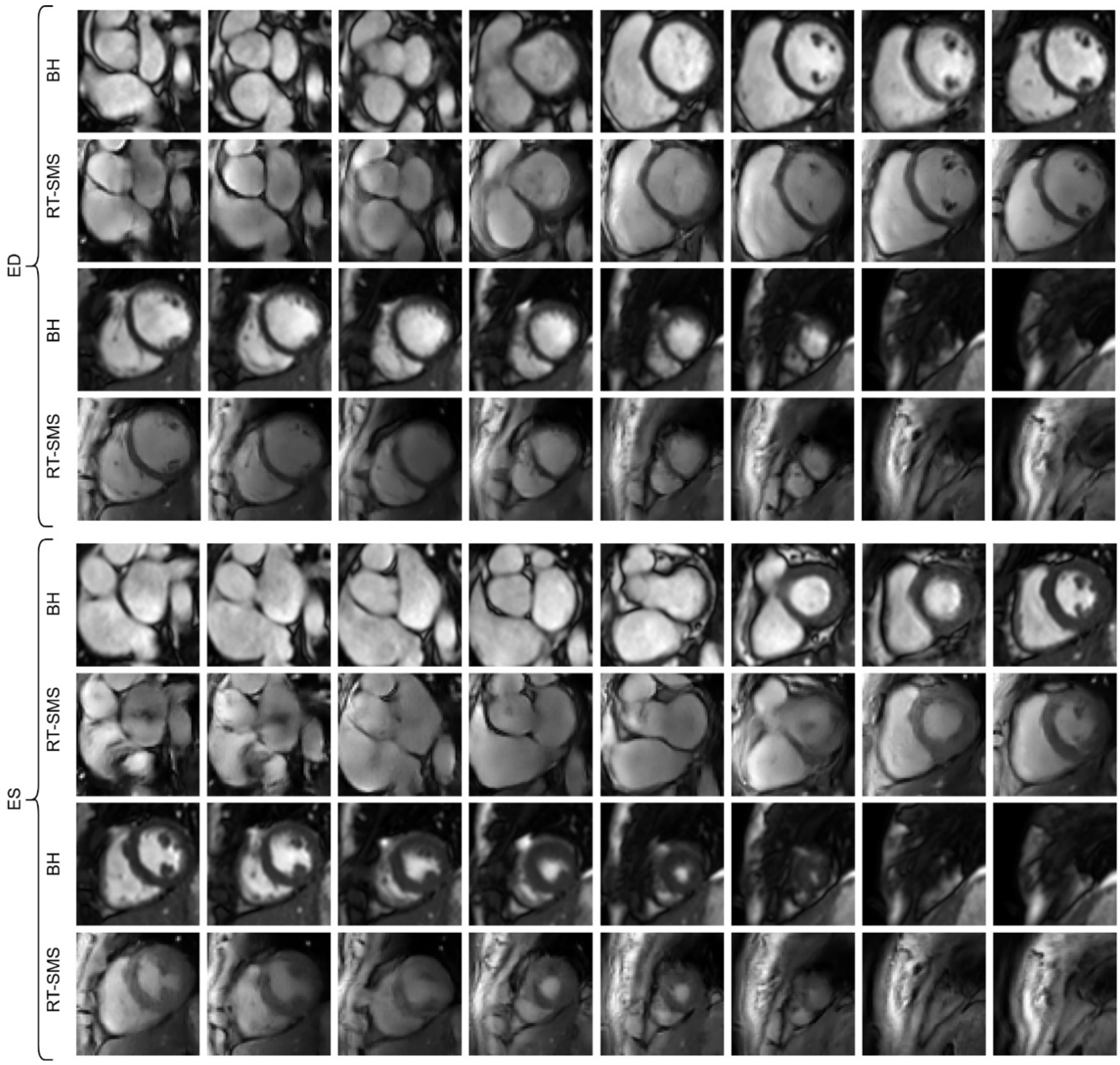

**Figure 6**: *Comparison between the RT-SMS acquisition with deep artifact suppression reconstruction and the reference-standard breath-hold (BH) acquisition across 16 concomitant slices in end diastole (ED) and end systole (ES) from a single subject.*

*Quantitative and Qualitative Image Quality*

Quantitative image quality metrics are displayed in Table 2. For RT-SMS data, deep artifact suppression significantly outperformed CS reconstruction in terms of edge sharpness, eSNR, and bmCR (all $p<0.001$). However, reference-standard BH imaging outperformed both RT-SMS reconstructions, across all metrics (all $p<0.001$).

| **Data Type** | **Edge sharpness ($mm^{-1}$)** | **eSNR** | **bmCR** |
|---|---|---|---|
| RT-SMS-DL | 0.18 ± 0.03 | 67.3 ± 53.9 | 2.21 ± 0.26 |
| RT-SMS-CS | 0.15 ± 0.03 | 36.3 ± 21.5 | 1.97 ± 0.19 |
| Reference-standard BH | 0.23 ± 0.01 | 195.6 ± 97.1 | 3.60 ± 0.16 |

**Table 2:** *Quantitative image quality metrics for RT-SMS reconstructed with deep artifact suppression (RT-SMS-DL) and CS (RT-SMS-CS), and for reference-standard BH images. Values are mean ± SD from two slices across 10 subjects. Repeated measures ANOVA demonstrated a significant effect of acquisition type for all metrics (all $p<0.001$) and Holm-corrected paired t-tests showed significant differences between all acquisition types for each metric (all $p < 0.001$).*

Qualitative ranking showed that RT-SMS-DL was preferred over RT-SMS-CS in 39/40 cases. Likert image quality scoring indicated that RT-SMS-DL provided adequate (or better) diagnostic image quality in 85% of datasets. Nevertheless, reference-standard BH imaging achieved higher diagnostic quality scores overall (median 5.0 [IQR 4.5–5.0] vs 3.0 [IQR 3.0–3.0], respectively, $p<0.001$).

*Volumetric Assessment*

Table 3 summarizes LV and RV volumetric measurements, with Bland-Altman analysis shown in Figure 7 and Figure 8. RT-SMS-DL demonstrated good agreement with reference-standard BH measurements, with small mean biases and narrow limits of agreement. Agreement was stronger for LV metrics than RV metrics.

| | **Reference-standard BH** | **RT-SMS-DL** | **Bias (LoA)** | **Correlation (r)** |
|---|---|---|---|---|
| **LV** | | | | |
| EDV (ml) | 162.4 ± 28.4 | 154.9 ± 28.3* | -7.5 (-20.8 to 5.8) | 0.97 |
| ESV (ml) | 59.4 ± 15.4 | 58.6 ± 12.7 | -0.8 (-9.0 to 7.3) | 0.97 |
| SV (ml) | 102.9 ± 16.4 | 96.3 ± 18.2* | -6.6 (-20.5 to 7.3) | 0.92 |
| EF (%) | 63.8 ± 5.7 | 62.3 ± 4.3 | -1.5 (-7.3 to 4.3) | 0.86 |
| Mass | 110.4 ± 22.2 | 100.0 ± 16.5* | -10.3 (-31.9 to 11.3) | 0.88 |
| **RV** | | | | |
| EDV (ml) | 187.6 ± 32.0 | 181.2 ± 35.8* | -6.4 (-22.9 to 10.1) | 0.98 |
| ESV (ml) | 84.6 ± 22.7 | 84.8 ± 24.1 | 0.2 (-20.7 to 21.2) | 0.90 |
| SV (ml) | 103.0 ± 16.2 | 96.3 ± 15.1 | -6.7 (-27.3 to 14.0) | 0.78 |
| EF (%) | 55.4 ± 7.1 | 53.8 ± 6.3 | -1.6 (-12.2 to 8.9) | 0.69 |

***Table 2:*** *Results from the volumetric measurements in reference-standard breath-hold (BH) and RT-SMS data with deep artifact suppression reconstruction (RT-SMS-DL). The table shows the mean ± standard deviation of the measurements, the bias and limits of agreement (LoA) between the measurements, as well as the corresponding correlation. * indicates a statistically significant difference between BH and RT-SMS-DL measurements (paired t-test, $p < 0.05$).*

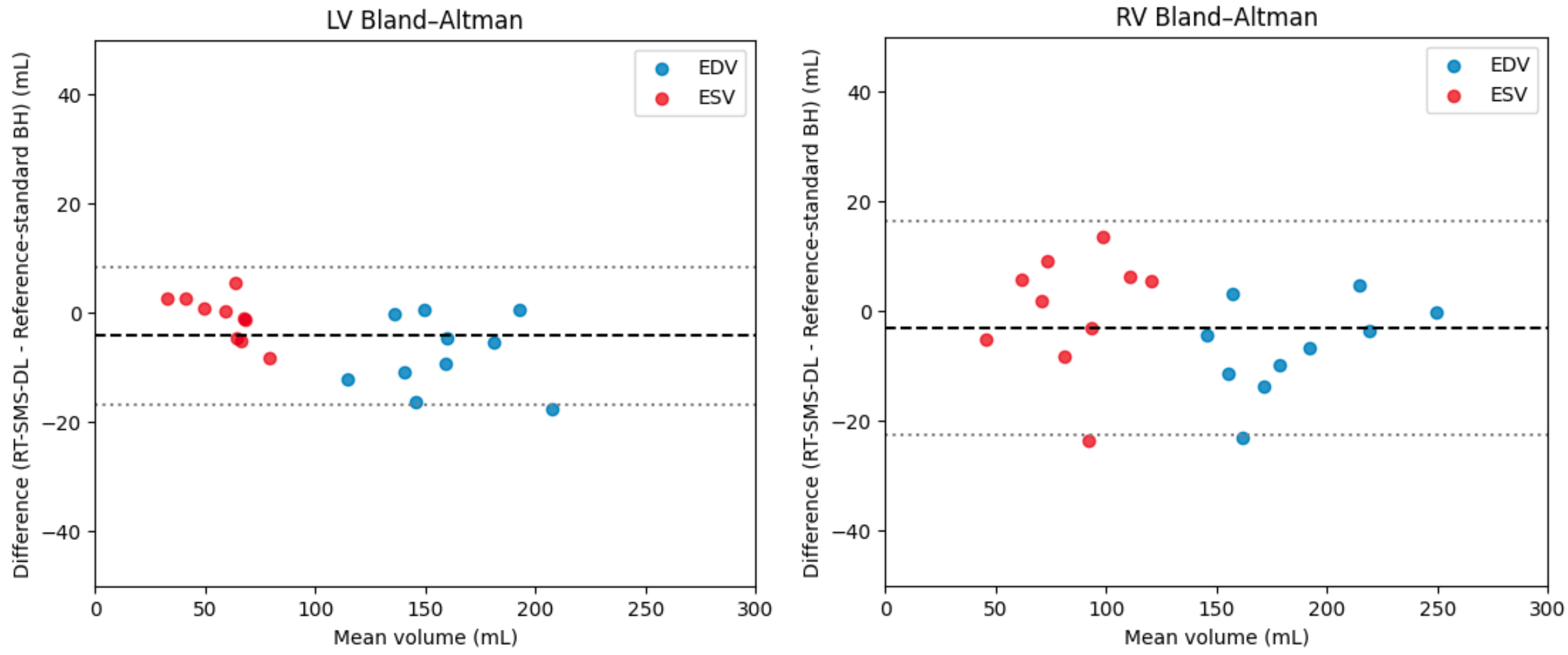

***Figure 7:*** *Bland-Altman plots showing the agreement in LV (left) and RV (right) volumes between RT-SMS-DL and reference-standard BH imaging, for end-diastolic volume (EDV, blue) and end-systolic volume (ESV, red).*

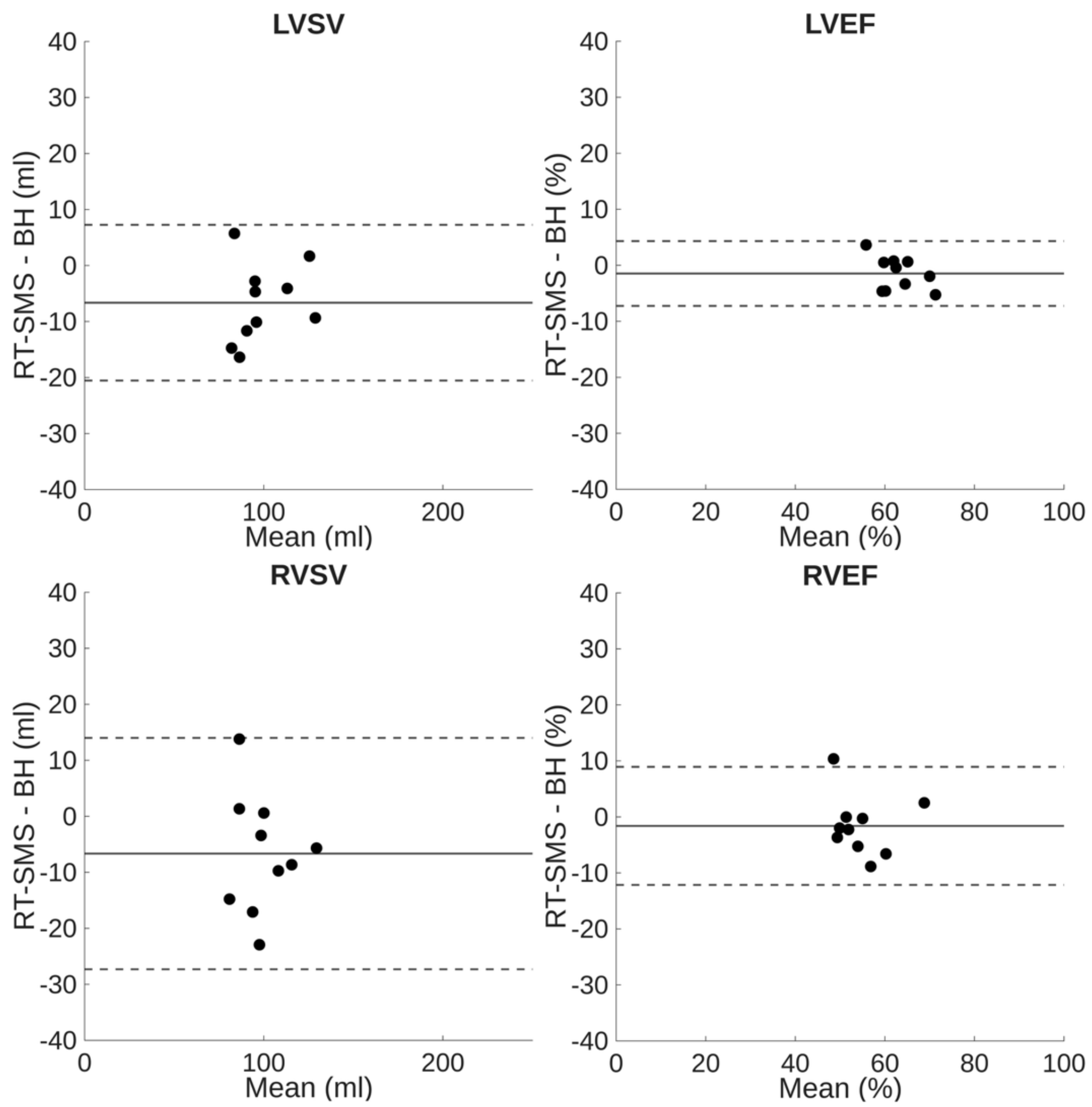

***Figure 8:*** *Bland-Altman plots showing the agreement in left-ventricular (LV) and right-ventricular (RV) stroke volume (LVSV and RVSV) and ejection fraction (LVEF and RVEF) between RT-SMS and reference-standard breath-hold (BH) imaging.*

## DISCUSSION

This study demonstrates the feasibility of rapid real-time CMR using a non-Cartesian RT-SMS acquisition, combined with online deep artifact suppression reconstruction. The proposed approach enables complete short-axis ventricular coverage within a single free-breathing acquisition of ~15 s, with reconstruction latency of ~30 s (~50× faster than compressed sensing). This represents a significant step towards the routine clinical adoption of fully real-time, free-breathing volumetric CMR using SMS. This is particularly relevant for patients with arrhythmia, and those who are unable to perform repeated breath-holds, where conventional cine imaging is often challenging or non-diagnostic.

Compared to state-of-the-art compressed sensing reconstruction of RT-SMS, deep artifact suppression provided substantially improved image quality, including sharper edges, higher eSNR and improved temporal fidelity. This is likely due to the chosen variable density spiral trajectory, which was optimized for DL-based reconstruction, but may therefore be disadvantageous to CS.

Subjective image quality for the RT-SMS acquisition with our DL-based reconstruction pipeline was rated as at least *adequate* in the majority of cases, and no cases were rated as non-diagnostic. In addition, ventricular volumetric measurements showed good agreement with reference-standard BH cine imaging, with low bias in calculated volumes. The better agreement in LV metrics compared to RV metrics could be due to the greater difficulty associated with RV segmentations, resulting in higher intra-observer variability. In addition, in prospective RT-SMS data, residual fat blurring was observed, attributable to off-resonance sensitivity of the long spiral readouts. This could be mitigated using gradient input response function (GIRF) correction[23] or other off-resonance compensation strategies, as demonstrated in previous work[5,24].

In contrast to many prior DL-based CMR reconstruction approaches[25], the proposed method operates directly on multi-coil complex image data. This preserves spatial encoding information that may otherwise be lost during coil combination, and this likely contributed to the improved artifact suppression and temporal consistency observed in preliminary experiments.

The proposed sequence is conceptually similar to prior SMS spiral bSSFP implementations at 0.55 T[24], however, first-order moment nulling of the readout gradient was not employed. In preliminary experiments, first-order nulling did not provide observable benefits at 1.5 T and instead exacerbated banding artifacts due to the associated increase in TR[26,27]. By omitting first-order nulling, a shorter TR (4.5 ms) was achieved, enabling a higher temporal resolution (36 ms) than previously reported for real-time SMS at 0.55 T (45 ms)[5].

More broadly, the presented work extends previous real-time and DL-based CMR approaches by combining non-Cartesian spiral sampling, SMS acceleration, and low-latency online DL reconstruction within a single pipeline. To our knowledge, this is the first demonstration of online deep learning reconstruction for non-Cartesian real-time SMS cine imaging at 1.5 T with full ventricular coverage.

**Limitations**

This study has several limitations. First, the sample size was small and limited to healthy volunteers; further validation in larger and pathological cohorts is required. In particular, performance in patients with arrhythmia or irregular breathing patterns, where real-time imaging is most clinically valuable, remains to be established. Second, only an SMS factor of two was investigated; higher SMS factors may provide further acceleration but may also increase slice leakage and require retraining of the DL model. Third, all data were acquired on a single scanner platform and field strength, and generalisation to other systems remains to be established. Fourth, the DL model was trained on retrospectively simulated data, which may not fully capture all artifacts present in prospectively acquired RT-SMS data. Future work could incorporate prospectively acquired training data or domain adaptation strategies to further improve robustness. Finally, no comparison was made with alternative DL reconstruction strategies (e.g. unrolled networks), which may offer different trade-offs between quality and speed.

Future work will focus on validation in larger and pathological cohorts, exploration of higher SMS acceleration factors, and integration of physics-based corrections such as GIRF and off-resonance compensation within the DL reconstruction framework. Ultimately, prospective clinical studies will be required to determine whether RT-SMS imaging can replace conventional breath-hold cine imaging in routine clinical workflows.

## CONCLUSION

Rapid free-breathing CMR with online reconstruction is feasible using non-Cartesian RT-SMS bSSFP combined with deep artifact suppression. The proposed method enables full short-axis ventricular coverage within a single ~15 s acquisition and <30 s reconstruction, while providing clinically acceptable ventricular volumetric measurements. This approach substantially reduces both acquisition and reconstruction time and supports the translation of fully real-time, free-breathing cine CMR into routine clinical practice, particularly for patients unable to perform breath-holds.

## CONFLICT OF INTEREST

Iulius Dragonu is an employee of Siemens Healthcare.